# Multiferroic Properties of $BaFe_{12}O_{19}$ Ceramics

*Xiuna Chen, Guolong Tan* *

*State Key Laboratory of Advanced Technology for Materials Synthesis and Processing,
Wuhan University of Technology, Wuhan 430070, China*

**Abstract**

Simultaneous occurrence of big ferroelectricity and strong ferromagnetism has been observed in Barium Hexaferrite ceramics. Barium hexaferrite ($BaFe_{12}O_{19}$) powders with hexagonal crystal structure have been successfully synthesized by polymer precursor method using barium acetate and ferric acetylacetonate as the precursor. The powders were pressed into pellets, which were sintered into ceramics at 1100°C~1300°C for 1 hour. The structure and morphology have been determined by X-ray diffraction (XRD) and field emission scanning electron microscopy (FESEM). Large spontaneous polarization was observed in the $BaFe_{12}O_{19}$ ceramics at room temperature, exhibiting a clear ferroelectric hysteresis loop. The maximum remanent polarization of $BaFe_{12}O_{19}$ ceramic is estimated to be Pr~11.8 $\mu C/cm^2$. The $FeO_6$ octahedron in its perovskite-like hexagonal unit cell as well as the shift of $Fe^{3+}$ off the center of octahedron are proposed to be the origin of polarization in $BaFe_{12}O_{19}$. The $BaFe_{12}O_{19}$ ceramic also shows strong ferromagnetism at room temperature.

**Keywords:** $BaFe_{12}O_{19}$, Multiferroic, Ferroelectric Properties, Ferromagnetic properties, Functional applications.

## Ⅰ Introduction

In recent years, there has been increasing interest in multiferroic materials, they provide a wide range of potential applications such as multiple-state memory elements, novel memory media, transducers and new functional sensors[1, 2]. However, the materials in which ferroelectricity and ferromagnetism coexist are rare[3, 4] and mostly exhibit rather weak ferromagnetism. Because the room-temperature multiferroism is essential to the realization of multiferroic devices that exploit the coupling between ferroelectric and ferromagnetic orders at ambient conditions, BiFeO$_3$ together with more recently revealed LuFe$_2$O$_4$, Pb$_2$Fe$_2$O$_5$ and PbFe$_{12}$O$_{19}$ [5, 6, 7, 8, 9] are currently considered to be promising candidates for practical device applications. The perovskite BiFeO$_3$ exhibits weak magnetism, which could somehow prevent its practical application. Therefore preparation of a material in which large ferroelectricity and strong ferromagnetism coexist would be a milestone for modern electrics and functionalized materials[10]. As large ferroelectric polarization was found in PbFe$_{12}$O$_{19}$ ceramics[9] with hexagonal structure, it opens up a new direction for potential multiferroic candidate in such traditional ferromagnetic oxides as BaFe$_{12}$O$_{19}$ which holds similar perovskite-like lattice units in its hexagonal structure.

* Corresponding author; Tel: +86-27-87870271; fax: +86-27-87879468. Email address: gltan@whut.edu.cn

M-type hexaferrites denoted as BaFe$_{12}$O$_{19}$ has attracted a lot of attention because of their excellent magnetic properties and potential application in various fields.[11] As a hexaferrite, BaFe$_{12}$O$_{19}$ is one of the mostly used ferrites in applications as permanent magnets[12]. The magnetization per formula unit at 0K is $(8-4)\times5=20\mu B$.[13] Due to its excellent magnetic properties, BaFe$_{12}$O$_{19}$ could hold a great promise to be a good applicable Pb-free multiferroic candidate in case suitable ferroelectricity could be observed.

Magnetic field induced electric polarization has been observed in BaFe$_{12-x-\delta}$Sc$_x$Mg$_\delta$O$_{19}$ ($\delta=0.05$) at 10K[14]. However, the multiferroic aspect of pure BaFe$_{12}$O$_{19}$ has never been studied yet. In this paper we are going to present large multiferroic effect in BaFe$_{12}$O$_{19}$ compound. The fabrication of BaFe$_{12}$O$_{19}$ powders by a polymer precursor method as well as the large spontaneous polarization of BaFe$_{12}$O$_{19}$ ceramic in addition to its strong ferromagnetism at room temperature will also be presented.

## Ⅱ Experimental procedure

Barium hexaferrite "BaFe$_{12}$O$_{19}$" powders were prepared by polymer precursor method using barium acetate (Ba(CH$_3$COO)$_2$) and ferric acetylacetonate as the starting materials. Typically, 0.2580 g barium acetate was dissolved in 15 mL distilled water to form a clear solution. The solution was stored in a three neck glass bottle. In order to prevent the hydrolysis of ferric acetylacetonate, the following experiments were carried out in the glove box. 3.7082g ferric acetylacetonate was dissolved in 200mL benzene. The prepared barium acetate solution and the ferric acetylacetonate in benzene, under the stoichiometric atomic ratio of Ba/Fe=1/10.5 was continuously heated and stirred in order to induce a homogenous mixture of both solutions at 50℃ for 1 hour. Then, 100ml ammonia and 15ml polyethylene glycol mixture solution was added into the above solution, thus a colloid dispersion solution was formed. The dispersion solution was maintained at 323K with stirring for 8 hours. Afterwards, the colloid solution was moved out of the glove box. The water and organic solution were removed by centrifugation. the remaining colloid powders were calcined at 450℃ for 1.5h to completely remove the organic part. 0.10g powders were weighted and pressed into pellet, which was then sintered into ceramics at 1200℃ and 1300℃ for 1 hour, respectively. Phase identification was performed by X-ray powder diffraction (XRD) method with Cu $K_\alpha$ radiation. The images for the morphology and microstructure of the ceramics were collected by S-4800 field emission scanning electron microscopy. All the sintered samples should be polished and both surfaces of samples were coated with silver paste as electrodes for P-E hysteresis loop measurement. The ferroelectric hysteresis loop was measured using a home-made instrument, termed as ZT-IA ferroelectric measurement system. The magnetization of the samples was measured using a Quantum Design physical property measurement system (PPMS).

## Ⅲ Results and discussion

### (1). Structure and microstructure of $BaFe_{12}O_{19}$ ceramics

The XRD patterns of the $BaFe_{12}O_{19}$ powders being calcined at 1200°C and 1300°C, respectively, are shown in *Figure 1*, which showed the hexagonal structure of pure $BaFe_{12}O_{19}$ in single phase. The XRD patterns of $BaFe_{12}O_{19}$ powders demonstrate the magnetoplumbite structure with no extra reflections and is perfectly indexed to (110), (112), (107), (114), (200), (203) and (2,0,11) crystal plane of hexagonal $BaFe_{12}O_{19}$ (PDF# 43-0002). The relative intensities of (107) and (114) peaks, which correspond to the inclined c-axis orientation, are higher than those of (112) and (200) peaks. These results indicate that the grains of $BaFe_{12}O_{19}$ ceramics are randomly oriented.

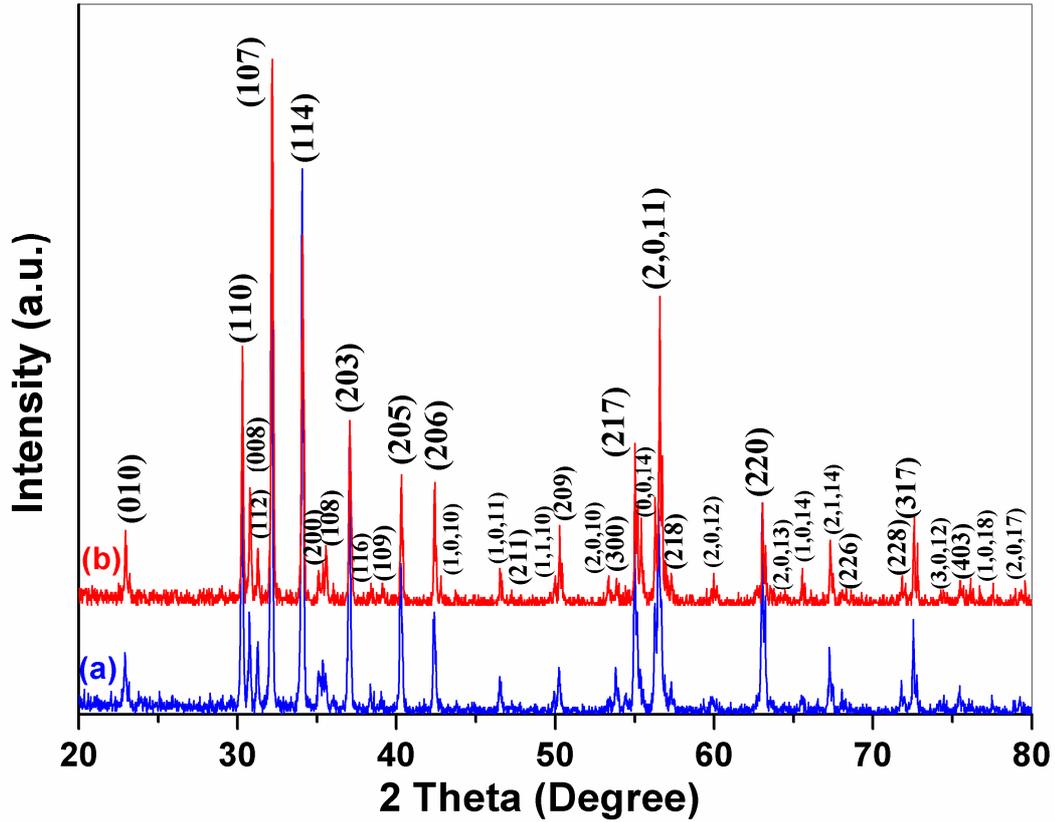

Figure 1: XRD pattern of the BaFe$_{12}$O$_{19}$ powders being calcined at (a) 1200℃ and (b) 1300℃, respectively.

*Figure 2* (a) and (b) show two FESEM images of BaFe$_{12}$O$_{19}$ ceramic being sintered at 1200℃ for one hour. The image clearly shows that the BaFe$_{12}$O$_{19}$ grains appear the hexagonal plate-like shape with the random orientation, and the grain size of BaFe$_{12}$O$_{19}$ ceramic is estimated to be less than 1μm. *Figure 1* (c) and (d) exhibit two FESEM images for the BaFe$_{12}$O$_{19}$ ceramic being sintered at 1300℃ for one hour. It can be seen that BaFe$_{12}$O$_{19}$ grains take plate-like shape, the grain size of which is among the range of 2~10 μm. Its densification has much more improved in comparison with that being sintered at 1200℃. Therefore, with the increase of sintering temperature, there is a great growth in grain size and improvement in densification for the BaFe$_{12}$O$_{19}$ ceramics. The distorted hexagonal flaky grains are frequently observed in the *BaFe$_{12}$O$_{19}$* ceramics, as being shown in *Figure 2*. Cylinder shaped grains can also be found in the SEM images. EDX measurement indicates that Ba rich phase has deposited onto the grain boundary area in BaFe$_{12}$O$_{19}$ ceramics through liquid phase sintering process, which expedites the rapid growth of the grain size of the ceramics. The hexahedron shaped grains are consistent with the hexagonal symmetry of BFO crystal. [0001] direction is normal to the surface of the flaky grains. The flaky shape suggests that the grains did not take priority growth along with [0001] direction but with [0100] direction. Obviously the growth rate of the grains along with [0100] direction is much faster than that along with other directions. Thus the grains did not take perfect hexahedron shapes, but actually the deformed flaky ones.

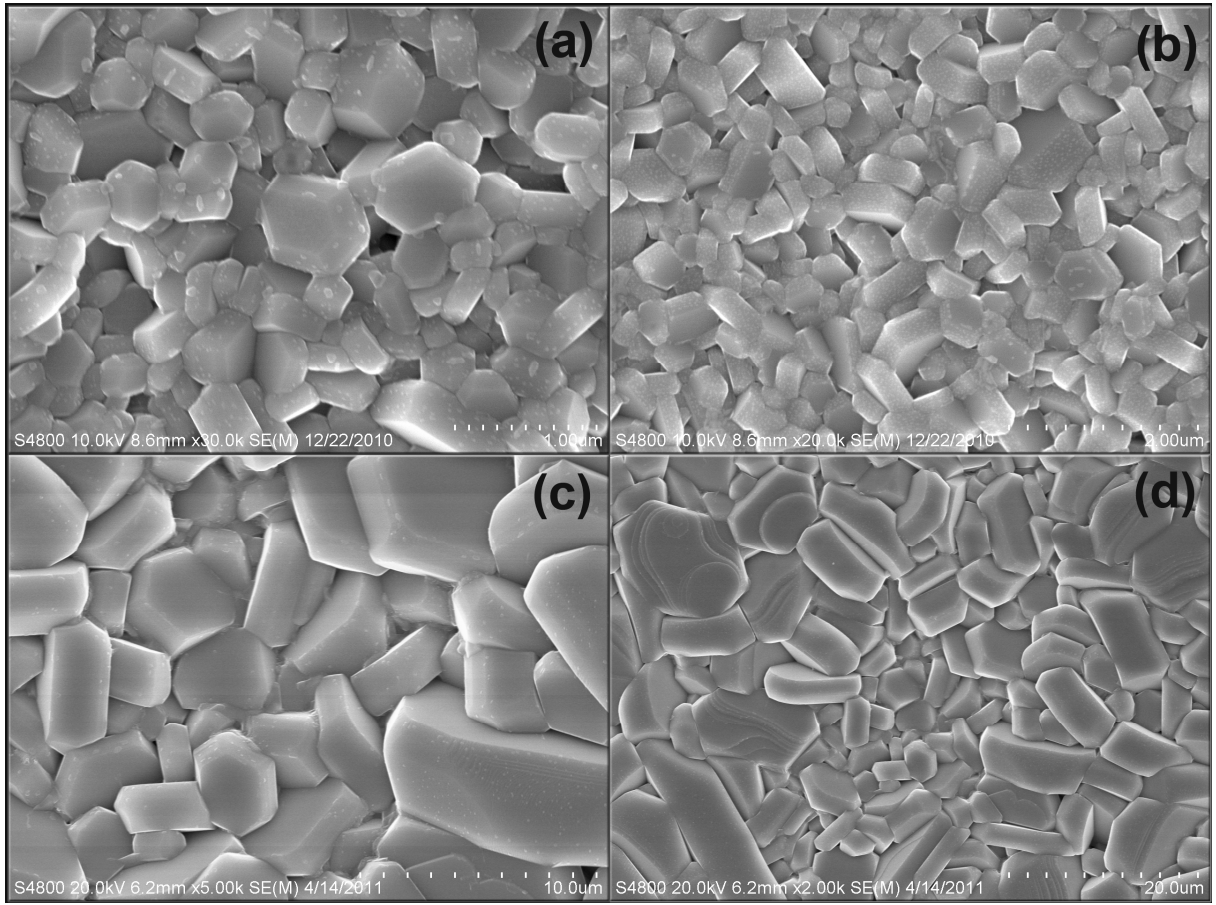

Figure 2: FESEM images of BaFe$_{12}$O$_{19}$ ceramics being sintered at (a), (b) 1200 ℃ and (c), (d) 1300 ℃ for 1h.

*(2) Ferroelectric properties of BaFe$_{12}$O$_{19}$ ceramics*

Ferroelectric properties were characterized using polarization hysteresis and pulse polarization measurements. The electric field-induced polarization behavior was examined using a home-made ferroelectric measurement system termed as ZT-IA. During the ferroelectric measurement, the specimen was parallel connected with a capacitor of 0.1μF for compensation. The F-E measurement was carried out by a tri-angular wave voltage. Evidence for the characterization of ferroelectric state of *BaFe$_{12}$O$_{19}$* ceramic is provided in *Figure 3*, which shows polarization cycles exhibiting clear ferroelectric hysteresis loops in BaFe$_{12}$O$_{19}$ ceramics under applied electric fields of different amplitudes obtained at room temperature. The maximum remanent polarization ($P_r$) and the coercive electric field ($E_c$) obtained from the ferroelectric hysteresis loop in *Figure 3* (a) are ~ 11.8μC/cm$^2$ and ~ 5.8kV/m, respectively, for the BaFe$_{12}$O$_{19}$ ceramic being sintered at 1200℃ for 1 hour.

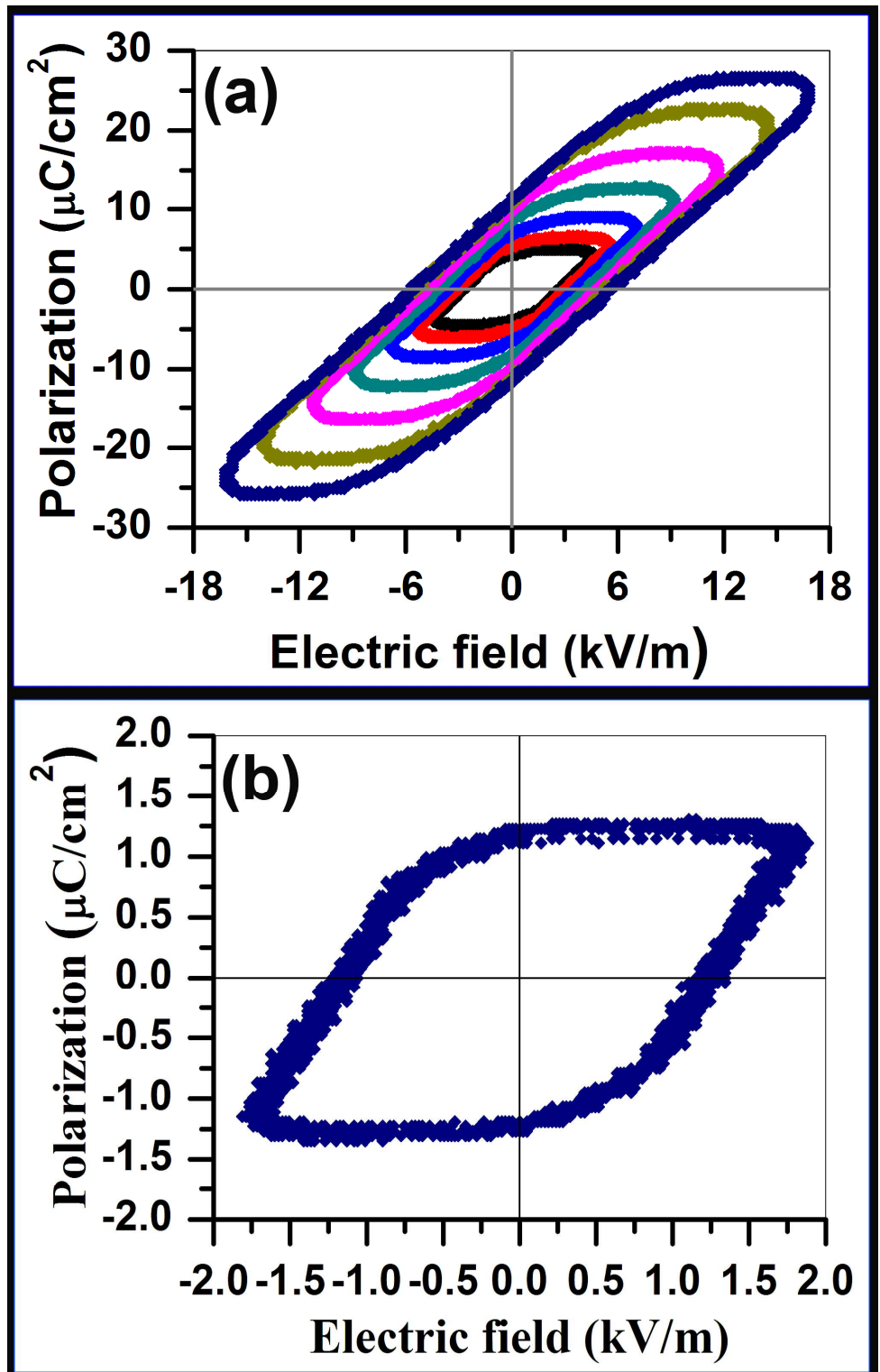

*Figure 3*: Ferroelectric hysteresis loops of BaFe$_{12}$O$_{19}$ ceramic being sintered at (a) 1200 ℃ and (b) 1300 ℃ for 1h, respectively.

After BaFe$_{12}$O$_{19}$ ceramic has been sintered at 1300℃ for 1 hour, the grain size and densification has been greatly improved in comparison with that being sintered at 1200 ℃, as being seen from their SEM images in *Figure 2*. The polarization hysteresis loop for the BaFe$_{12}$O$_{19}$ ceramic being sintered at 1300℃ exhibits a F-E loop being closer to the standard one, which saturates at a certain field and demonstrates some convex and concave regions in the curve, as being shown in *Figure 3* (b). The saturated polarization of the BaFe$_{12}$O$_{19}$ ceramics being sintered at 1300℃ depmonstrates a quick reduction with the increase of the densification and grain size. The maximum remanent polarization (P$_r$), the polarization maximum (P$_{max}$) and the coercive electric field (E$_c$) obtained from the ferroelectric hysteresis loop in *Figure 3* (b) are ~1.2μC/cm$^2$, ~1.6μC/cm$^2$ and ~1.25kV/m respectively for BaFe$_{12}$O$_{19}$ ceramic pellet being sintered at 1300℃. Conclusion could be drawn from *Figure 3* (a) & (b) that the ferroelectricity is reduced with the increase of the sintering temperature due to the improvement of densification and growth of the grain size of the ceramics.

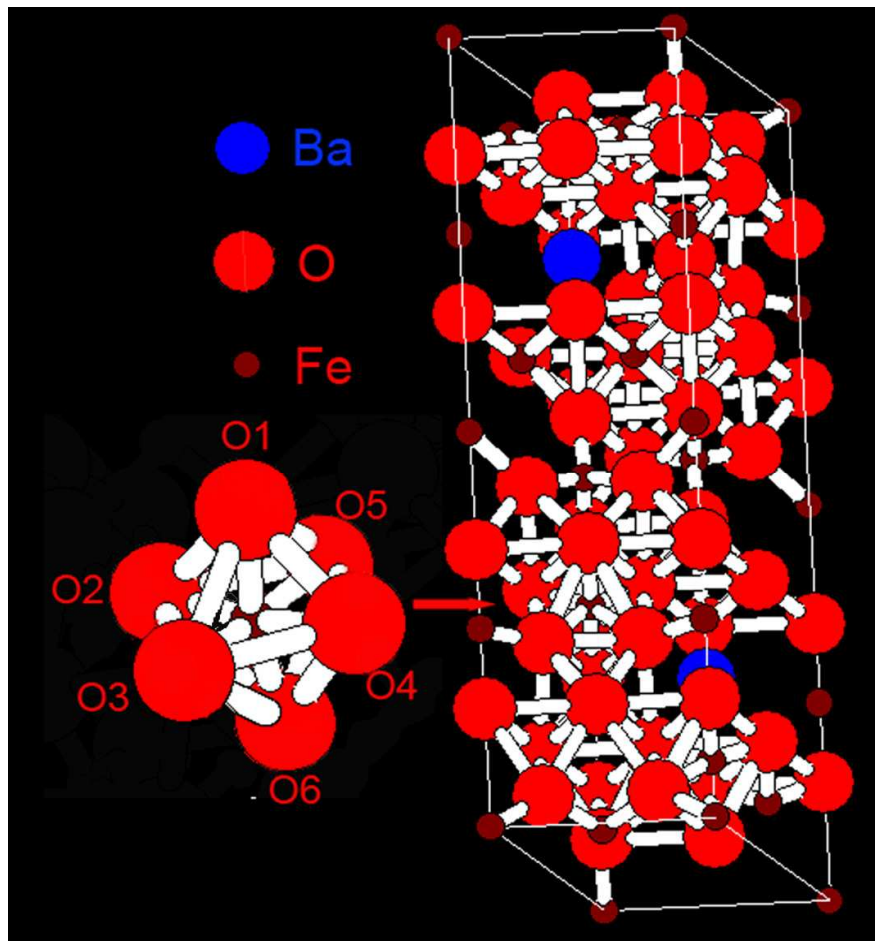

*Figure 4: Crystal structure model of BaFe$_{12}$O$_{19}$ (ICSD #45442) and magnified FeO$_6$ octahedron perovskite structure found in the crystal structure model of BaFe$_{12}$O$_{19}$.*

In order to understand the origin of the ferroelectricity of barium hexaferrite, we firstly investigated the crystal structure model of $BaFe_{12}O_{19}$ (ICSD #45442) with space group P63/mmc, which is exhibited in *Figure 4*. Careful analysis of the unit model structure suggests a perovskite-like crystal structure with one distorted $FeO_6$ oxygen octahedron in hexagonal $BaFe_{12}O_{19}$ as being shown in *Figure 4*. Each hexagonal $BaFe_{12}O_{19}$ model has one $FeO_6$ oxygen octahedron in a sub-unit cell. In a normal octahedron, Fe cation is located at the center of an octahedron of oxygen anions. However, in the unit cell of $BaFe_{12}O_{19}$ below the Curie temperature, there is also a distortion to a lower-symmetry phase accompanied by the shift off-center of the small Fe cation. Fe cation shifts away from the center along b axis, while $O_5$ and $O_6$ shifts off their original positions of octahedron along opposite directions of a-axis, which leads to the distortion of $O_5$-Fe-$O_6$ bond away from straight line. The spontaneous polarization derives largely from the electric dipole moment created by the two shifts, which induces the large ferroelectric hyetersis loops for $BaFe_{12}O_{19}$.

### *(3) Ferromagnetism of $BaFe_{12}O_{19}$ ceramics*

$BaFe_{12}O_{19}$ is a kind of traditional ferromagnetic material, whose magnetization behavior has been widely studied. Magnetic properties of the $BaFe_{12}O_{19}$ sample being sintered at 1300°C were measured at room temperature with a Quantum Design physical property measurement system (PPMS). The field-dependent magnetization hysteresis loop for $BaFe_{12}O_{19}$ sample being sintered at 1300°C is shown in *Figure 5*. The remnant magnetic polarization ($M_r$) of the $BaFe_{12}O_{19}$ is 32 emu/g, the saturation magnetization is ~55 emu/g. The coercivity ($H_c$) of the $BaFe_{12}O_{19}$ sample is 1607 Oe. The magnetization of our $BaFe_{12}O_{19}$ sample is lower than the bulk value of 67.7 emu/g. Meanwhile, The $BaFe_{12}O_{19}$ sample being sintered at 1200°C exhibit reduced magnetization value in comparison with its counterpart being sintered at higher temperature (1300°C). According to the SEM observation, the grain size of the $BaFe_{12}O_{19}$ ceramic being sintered at 1200°C was less than 1μm, while that being sintered at 1300°C was estimated to be among the range of 2~10μm. The particle size of $BaFe_{12}O_{19}$ sample grew very fast with sintering temperature. The reduction of the magnetization for $BaFe_{12}O_{19}$ samples with decreasing particle size was caused by the incomplete coordination of the atoms on the particle surface, leading to a noncollinear spin configuration, which causes the formation of a surface spin canting[15, 16, 17], and due to thermal fluctuation of magnetic moments, which significantly diminishes the total magnetic moment for a given magnetic field[15].

Size effect plays an important role on the reduction of the magnetization of $BaFe_{12}O_{19}$ ceramics. Usually the so-called critical size determines the properties of magnetic materials[18, 19, 20]. When particle size is smaller than this critical value, particles locate within one single domain; otherwise multiple-domain may occur in particles. As particles are larger than the single domain size for the ceramics being sintered at higher temperature, the domain walls become predominant. While the sintering temperature rises, the particle size increased towards to the critical single domain size, the coercivity increase and reach a maximum value at the single domain size because of the coherent rotation of spins.

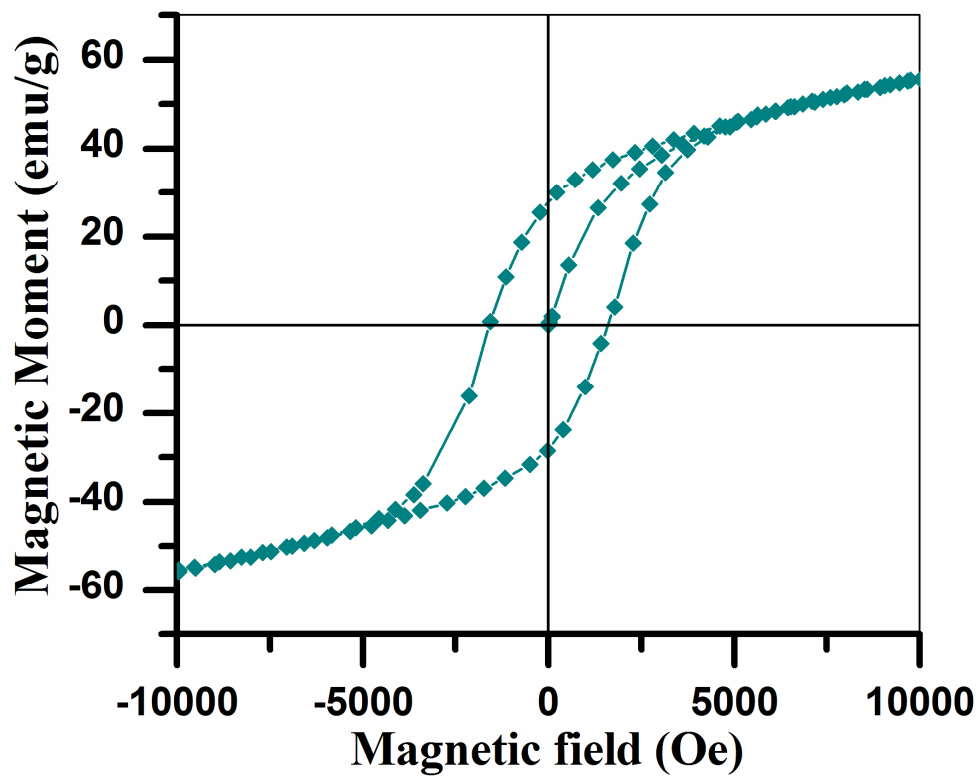

*Figure 5   Magnetic hysteresis loop of the $BaFe_{12}O_{19}$ ceramics being sintered at 1300 ℃.*

So far, $BiFeO_3$ is the best-known multiferroic material, which also exhibits both ferromagnetic and ferroelectric ordering above room temperature. However, the weak ferromagnetism of $BiFeO_3$ may prevent its application [21]. In contrast, strong ferromagnetism and large ferroelectricity simultaneously occurred in $BaFe_{12}O_{19}$ ceramics above room temperature. The remnant magnetic polarization ($M_r$) and the magnetic coercivity ($H_c$) of $BiFeO_3$ was reported to be less than 0.1 emu/g and 200 Oe[21], which are much less than that of ~32 emu/g and 1607 Oe for $BaFe_{12}O_{19}$ ceramics. The remnant magnetic polarization of $BaFe_{12}O_{19}$ ceramics is 320 times higher than that of $BiFeO_3$ ceramics, while the magnetic coercivity ($H_c$) is around 8 times higher than that of $BiFeO_3$ ceramics. Meanwhile, as being mentioned above, the maximum remnant polarization (Pr) of $BaFe_{12}O_{19}$ ceramics is determined to be ~32 μC/cm$^2$, which is around 5.2 times higher than the reported value of 6.1μC/cm$^2$ from $BiFeO_3$ ceramics[22]. Therefore multiferroic $BaFe_{12}O_{19}$ ceramics have clear and unique advantage over the best multiferroic system of $BiFeO_3$ materials. This holds promise for its application in new generation of electronic devices as a applicable multiferroic candidate in single phase.

## IV Conclusion

In summary, large ferroelectricity and strong ferromagnetism have been found in barium hexaferrite ceramics. pure barium hexaferrite ($BaFe_{12}O_{19}$) powders have been synthesized by a novel polymer precursor method using glycerin as solvent. The powders were pressed into pellets, which were sintered into ceramics at 1200℃ and 1300℃ for 1 hour, respectively. The ferroelectric hysteresis loop of the sample shows that the maximum remnant polarization ($P_r$) and the coercivity ($H_c$) are 11.8 μC/cm$^2$ and 5.8 kV/m, respectively for $BaFe_{12}O_{19}$ ceramics being sintered at 1200℃. We propose that the source of polarization is the distortion of the Fe oxygen octahedron in the lattice unit of its perovskite-like hexagonal structure. The magnetic hysteresis loop of the sample shows that the remnant magnetic polarization ($M_r$) and the magnetic coercivity ($H_c$) are ~11.8 emu/g and 1632 Oe, respectively. These results clearly demonstrate the multiferroic characterization of the barium hexaferrite ceramics above room temperature.

Acknowledgement: The authors greatly acknowledge the financial support from the Natural Science Foundation of Hubei Province (2010CDA078) and